\documentclass[aps,prc,twocolumn,superscriptaddress]{revtex4-2}
\usepackage{CJK}
\usepackage{graphicx}
\usepackage{hyperref}
\usepackage{amssymb}
\usepackage{amsmath}
\usepackage[normalem]{ulem}
\usepackage{url}
\usepackage{color}
\usepackage{multirow}
\usepackage{float}
\usepackage{xspace}
\usepackage{amsthm,amsmath,amssymb}
\usepackage{mathrsfs}
\usepackage{diagbox}
\usepackage[T1]{fontenc}

\begin{document}

	\title{Novel Pauli blocking method in quantum-molecular-dynamics-type models}
	
	\author{Xiang Chen}
	\affiliation{China Institute of Atomic Energy, P. O. Box 275(18), Beijing 102413, China}
	\author{Junping Yang}
	\affiliation{China Institute of Atomic Energy, P. O. Box 275(18), Beijing 102413, China}
	\author{Ying Cui}
	\email{yingcuid@163.com} 
	\affiliation{China Institute of Atomic Energy, P. O. Box 275(18), Beijing 102413, China}
	\author{Kai Zhao}
	\email{zhaokai$_$th@163.com} 
	\affiliation{China Institute of Atomic Energy, P. O. Box 275(18), Beijing 102413, China}
	\author{Zhuxia Li}
	\affiliation{China Institute of Atomic Energy, P. O. Box 275(18), Beijing 102413, China}

	\author{Yingxun Zhang}
	\email{zhyx@ciae.ac.cn}
	\affiliation{China Institute of Atomic Energy, P. O. Box 275(18), Beijing 102413, China}%
	\affiliation{Department of Physics and Technology, Guangxi Normal University, Guilin 540101, China}%

\date{\today}

\begin{abstract}

In this work, we propose a novel method for calculating the occupation probability in the Pauli blocking of the quantum molecular dynamics type models. This method refines the description of the Pauli blocking ratio in the nuclear matter and that in the finite nucleus. 
The influence of the new Pauli blocking method on the heavy ion collisions observables, such as the charge distribution, the free neutron to proton yield ratios, and the extracted physical quantities, such as the in-medium nucleon-nucleon cross sections, are investigated. For the extracted in-medium nucleon-nucleon cross sections, our results show that it will be enhanced 1.1$-$2.5 times than that with the conventional Pauli blocking method at the beam energy less than 150 MeV/u, which highlights the importance of a refined Pauli blocking method for developing an advanced transport model to describe complex heavy ion collisions. 

\end{abstract}

\maketitle
\textit{Introduction.}
Investigations of the medium properties of nuclear systems, such as the isospin asymmetric nuclear equation of state (EOS)~\cite{BALi2008PR,Russotto2023RNC,Sorensen2024,Lynch2022PLB} and in-medium nucleon-nucleon ($NN$) cross sections~\cite{YXZhang2007PRC,Henri2020PRC,PCLi2022PLB,Tian2023PRC}, play a crucial role in advancing our understanding of dense nuclear matter, dynamics of heavy ion collision (HIC), and effective nucleon-nucleon interaction. In the laboratory, HIC is a unique way to extract the information of interest about the EOS and in-medium $NN$ cross sections. However, these extractions rely on the transport models. As a game changer, developing advanced models have been suggested in the long-range plan for nuclear science~\cite{2023LRP}.

To develop advanced transport models, one has to understand the model dependence of the extraction of the EOS and in-medium $NN$ cross sections via HICs first. Since 2009, the transport model evaluation project (TMEP) has made important progresses in this area~\cite{JunXu2016PRC,YXZhang2018PRC,Ono2019PRC,Colonna2021PRC,Wolter2022PPNP}. The TMEP results illustrate that one of the crucial theoretical challenges for the transport model is the Pauli blocking. The commonly used Pauli blocking methods in the transport models underestimate the blocking ratio compared to its benchmark value, especially in the quantum molecular dynamics (QMD)-type models, in the nuclear matter~\cite{YXZhang2018PRC}. Thus, one may expect that the extracted in-medium $NN$ cross sections or EOS via HICs may have systematic deviations relative to the true value, and a refined Pauli blocking method is indispensable to develop advanced transport models in the future.

In the transport equation, the Pauli blocking comes from the Uehling-Uhlenbeck factor~\cite{Wolter2022PPNP,GJMao1996PRC,Danielewicz1984AP}, and is realized after $NN$ attempted collisions in transport models. For example, for the attempted collision between particles $i$ and $j$, the final states of the outgoing particles $i$ and $j$ are occupied by the surrounding nucleons with the probabilities $P_i$ and $P_j$, respectively. Then, the collision will be blocked according to the blocking probability $P=1-(1-P_i)(1-P_j)$. 

In QMD-type models, the calculations of the occupation probability $P_i$ can be generally divided into two kinds. One is to calculate $P_i$ from the phase space density~\cite{YXZhang2018PRC}, i.e.,
\begin{equation}\label{Eq:PBW}
\begin{aligned}
    P_i=P(\mathbf{r}_i,\mathbf{p}'_i) &=\frac{1}{4/h^3}\sum_{k=1,k\ne i}^{N} \frac{1}{(\pi\hbar)^3}\\
    & \mathrm{exp}\left[-\frac{(\mathbf{r}_i-\mathbf{R}_k)^2}{2\sigma_r^2}-\frac{(\mathbf{p}'_i-\mathbf{P}_k)^2}{2\sigma_p^2}\right], 
\end{aligned}
\end{equation}
where $\sigma_r$ and $\mathbf{R}_k$ ($\sigma_p$ and $\mathbf{P}_k$) are the width and centroid of the phase space density in coordinate (momentum) space. The $\mathbf{r}_i$ and $\mathbf{p}'_i$ are the coordinate and momentum of outgoing nucleon $i$. Another is to calculate $P_i$ from the overlap of the hard spheres~\cite{Aichelin1988PRC,Aichelin1991PR}, i.e.,
\begin{equation}
    P_i=P(\mathbf{r}_i,\mathbf{p}'_i)=\sum_{k=1,k\ne i}^{N} \frac{1}{4/h^3} O_{ik}^{(x)}O_{ik}^{(p)}.
\end{equation}
Here, $O_{ik}^{(x)}$ ($O_{ik}^{(p)}$) is the volume of the overlapping region of spheres with the parameters of radius $R_x$ ($R_p$) of nucleons $i$ and $k$ in coordinate (momentum) space. 
This method was developed as an approximation of Eq.(\ref{Eq:PBW}) since computing exponentials with computers was slow. The variant of the second method was to introduce the additional surface correction~\cite{Aichelin1991PR,Cozma2018EPJA}, since the spurious $NN$ collisions caused by conventional Pauli blocking method mainly occur in the surface region of finite nucleus~\cite{XChen2021CPC}. 


These two methods mentioned above cause a large fluctuation in the occupation probability $P_i$ which results in the underestimation of the average occupation probability, particularly in the QMD type models where each nucleon is represented by a fixed width of Gaussian wave packet or is in a specific state. To address this issue, a way to reduce the fluctuation of the occupation probability $P_i$ is required. 

In this work, we propose an \textit{ad hoc} method to calculate the occupation probability $P_i$ by considering the quantum effects on the distribution of the final states. Then, this new method is evaluated in the nuclear matter at $T$=5 MeV and in the finite nucleus used in the initialization of the transport model. 
To understand the influence of the new Pauli blocking method on the mechanism of heavy ion collisions, we select two observables. One is the charge distribution, which provides insights into the fragmentation pattern. Another is the spectra of the free neutron to proton yield ratios, which has been used to determine the density dependence of symmetry energy. Lastly, we investigate the influence of the new Pauli blocking method on the extraction of physical quantity, such as the in-medium $NN$ cross sections, by describing the stopping power data. 




\textit{Theoretical model.}
The improved quantum molecular dynamics (ImQMD) model we used is the same as in Refs.~\cite{YXZhang2014PLB,YXZhang2015PLB,YXZhang2020FP}, the nucleonic potential energy density $u$ without the spin-orbit term can be written as the sum of the local term $u_{\mathrm{loc}}$ and momentum dependent interaction term $u_{\mathrm{md}}$, i.e.,  $u=u_{\mathrm{loc}}+u_{\mathrm{md}}$. The local term is defined as 
\begin{equation}
\begin{split}
     u_{\mathrm{loc}} &= \frac{\alpha}{2} \frac{\rho^2}{\rho_0} +\frac{\beta}{\gamma+1} \frac{\rho^{\gamma+1}}{\rho_0^\gamma}+
     \frac{g_{\mathrm{sur}}}{2\rho_0}(\nabla \rho)^2\\
       &\quad +\frac{g_{\mathrm{sur,iso}}}{\rho_0}[\nabla(\rho_n-\rho_p)]^2+A_\mathrm{sym} \frac{\rho^2}{\rho_0}\delta^2+
     B_\mathrm{sym} \frac{\rho^{\gamma+1}}{\rho_0^\gamma}\delta^2.  
\end{split}
\end{equation}
Here, $\delta=(\rho_n-\rho_p)/(\rho_n+\rho_p)$ is the isospin asymmetry, $\rho_0$ is the saturation density, $\rho_n$ and $\rho_p$ are the densities of the neutron and proton, respectively. The coefficients of $\alpha$, $\beta$, $\gamma$, $g_{\mathrm{sur}}$, $g_{\mathrm{sur,iso}}$, $A_\mathrm{sym}$,  $B_\mathrm{sym}$ can be obtained from the standard Skyrme interaction parameters $t_0$, $t_1$, $t_2$, $t_3$, $x_0$, $x_1$, $x_2$, $x_3$, $\sigma $~\cite{YXZhang2006PRC}.
The momentum dependent interaction term can be derived from its interaction form $\delta(\mathbf{r}_1-\mathbf{r}_1) (\mathbf{p}_1-\mathbf{p}_2)^2$~\cite{Skyrme1956PhMa,Vautherin1972PRC,YXZhang2014PLB},
\begin{equation}
\begin{split}
     u_{\mathrm{md}} &= C_0 \sum_{ij}\int \mathrm{d}^3p\mathrm{d}^3p' f_i(\mathbf{r},\mathbf{p}) f_j(\mathbf{r},\mathbf{p'}) (\mathbf{p}-\mathbf{p'})^2        \\
       &\quad +D_0 \sum_{ij\in n}\int \mathrm{d}^3p\mathrm{d}^3p' f_i(\mathbf{r},\mathbf{p}) f_j(\mathbf{r},\mathbf{p'}) (\mathbf{p}-\mathbf{p'})^2     \\
       &\quad +D_0 \sum_{ij\in p}\int \mathrm{d}^3p\mathrm{d}^3p' f_i(\mathbf{r},\mathbf{p}) f_j(\mathbf{r},\mathbf{p'}) (\mathbf{p}-\mathbf{p'})^2,
\end{split}
\end{equation}
where $f_i(\mathbf{r},\mathbf{p})$ is phase space density distribution function of the $i$th nucleon. The parameters $C_0$ and $D_0$ can be determined from the standard Skyrme momentum dependent interaction term, and details can be found in Ref.~\cite{YXZhang2014PLB}. The treatment of initialization and $\emph{NN}$ collision used in this work are the same as those in Refs.~\cite{YXZhang2006PRC,YXZhang2007PRC}. 

The in-medium $NN$ elastic cross section $\sigma_{NN}^{\mathrm{med}}$ reflects the medium correction beyond the Pauli blocking, and is taken as
\begin{equation}
  \sigma_{NN}^{\mathrm{med}}=\left( 1+\eta(\sqrt{s})\frac{\rho}{\rho_0}\right)\sigma_{NN}^{\mathrm{free}}
\label{Eq:CroSec}
\end{equation}
in this work. The $\eta$ is a parameter that varies with incident energy, and $\sigma_{NN}^{\mathrm{free}}$ is free $NN$ elastic cross sections taken from Ref.~\cite{Cugnon1996NIM}. Equation (\ref{Eq:CroSec}) has been widely used in the transport models~\cite{YXZhang2007PRC,Basrak2016PRC,Persram2002PRC,Yuan2010PRC,Tsang2023arXiv}. 
Usually, the $\eta$ is determined by describing the HICs data with transport model simulations.

\textit{Pauli blocking method.}
The conventional Pauli blocking we used in ImQMD is as the same as Eq. (\ref{Eq:PBW}). To overcome the defect of the conventional Pauli blocking in QMD-type models, one possible approach is to consider the quantum effects on the states of nucleons, i.e., each nucleon can occupy a series of states that are labeled by the momentum. This effect has been considered in the antisymmetrized molecular dynamics (AMD) model~\cite{Ono1992PTP,Ono2004PPNP}, in which the wave function of a nuclear system is approximately described by a Slater determinant. 

\begin{figure}[htbp]
\includegraphics[angle=0,scale=0.35]{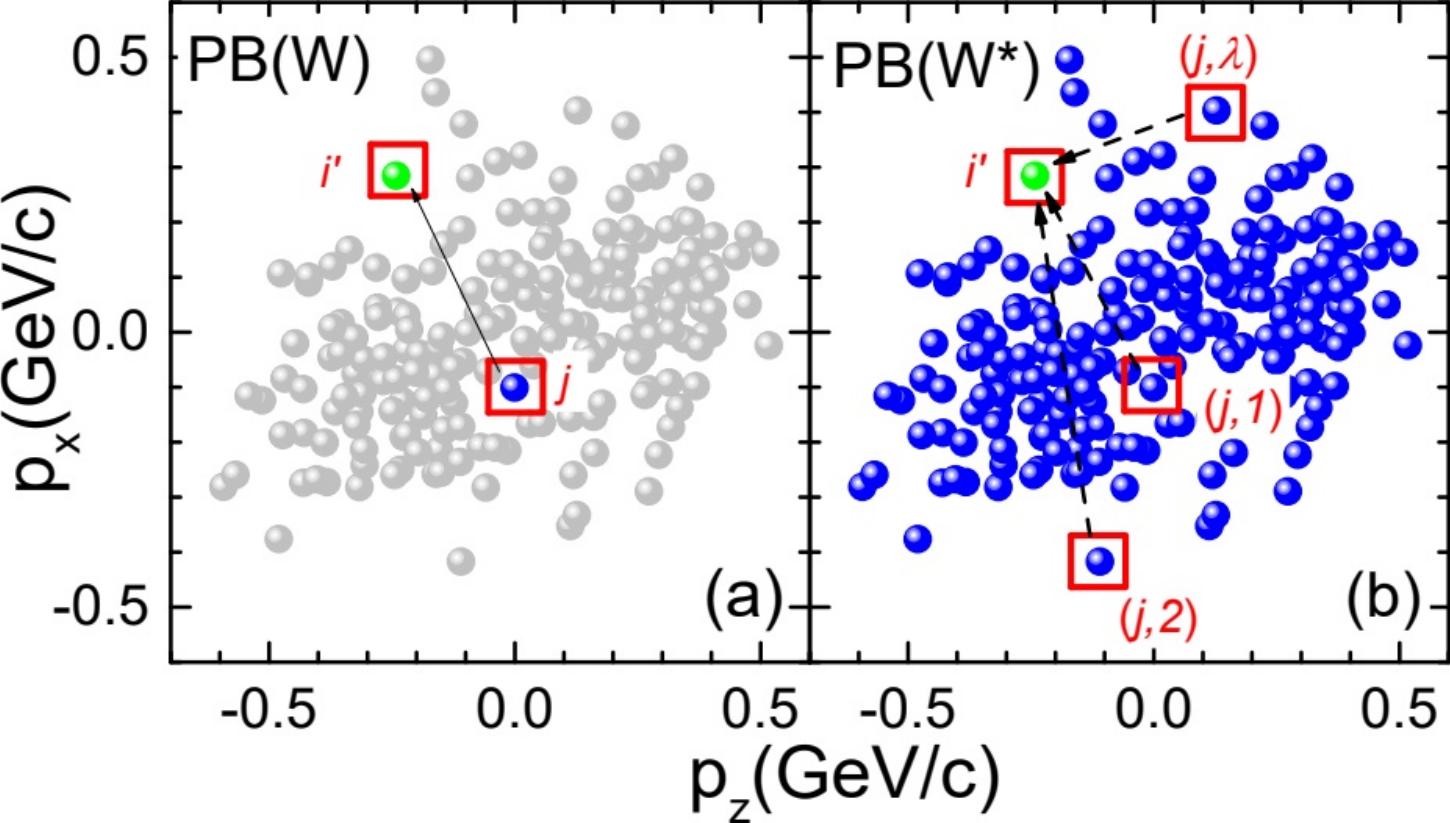}
\vspace{-1.0em}
\caption{Sketch of the idea for the conventional Pauli blocking method `PB(W)' and new Pauli blocking method `PB(W*)'.}
\label{fig:Fig1-Sketch}
\end{figure}

In QMD-type models, we propose an \textit{ad hoc} Pauli blocking method to calculate the occupation probability $P_i$. In this new method, the $P_i$ is also determined by the surrounding nucleons, but each nucleon has $N$ states. 
For understanding, we schematically present this idea in momentum space as in Fig.~\ref{fig:Fig1-Sketch}(b). The green point is the nucleon $i$ in its state $\mathbf{p}'_i$, and the blue points are nucleon $j$ in $N$ states labeled with $(j,\lambda=1, 2,...,N$). The corresponding formula of $P_i$ reads
\begin{equation}
\begin{aligned}
    P_i&=P(\mathbf{r}_i,\mathbf{p}_i') \\
    &= \frac{1}{4/h^3} \frac{1}{(\pi\hbar)^3} \sum_{j=1,j\ne i}^N \mathrm{exp} \left[-\frac{(\mathbf{r}_i-\mathbf{R}_{j})^2}{2\sigma_r^2}  \right]  c_j(\mathbf{p}'_i).
\end{aligned}
\label{Eq:OccPro}
\end{equation}
where the factor $c_j(\mathbf{p}'_i)$ is calculated as
\begin{equation}
\begin{split}
   c_j(\mathbf{p}'_i) = \frac{1}{N} \sum_{\lambda=1}^{N} \mathrm{exp} \left[-\frac{(\mathbf{p}'_i-\mathbf{P}_{j,\lambda})^2}{2\sigma_p^2}  \right].
\end{split}
\label{Eq:cpjl}
\end{equation}
$N$ is equal to the nucleon number of the system. 
The momentum $\mathbf{P}_{j,\lambda}$ represents the $j$th nucleon at state $\lambda$, and is sampled from the momentum space of the system. 
Thus, one can expect that $c_j(\mathbf{p}'_i)$ is smoother than that in the conventional method, where 
\begin{equation}
    c_j(\mathbf{p}'_i) = e^{-\frac{(\mathbf{p}'_i-\mathbf{P}_{j})^2}{2\sigma_p^2}},
\end{equation}
and the idea of the conventional method is schematically presented in panel (a) for comparison. In practical calculations, we approximate the Gaussian function in Eqs.(\ref{Eq:OccPro}) and (\ref{Eq:cpjl}) with a triangle to enhance the efficiency of the calculations. In the following discussions, we will refer to the new Pauli blocking method as PB(W*) and the conventional Pauli blocking as PB(W).

\textit{Occupation probability and successful collision rate in nuclear matter and the finite nucleus.}
To evaluate the new Pauli blocking method PB(W*), one has to establish benchmark values for comparison. Two benchmark values can be utilized for this purpose. The first one is the occupation probability $P(p')$ in nuclear matter. By comparing the calculated occupation probability with the Fermi distribution, we can evaluate the effectiveness of the Pauli blocking method in nuclear matter. The second benchmark is the occupation probability in the ground state of finite nuclei. This benchmark allows us to evaluate the performance of the new Pauli blocking method in reproducing the occupation probabilities in realistic nuclear systems. Deviations of the calculated occupation probabilities from the expected values provide insight into the accuracy and reliability of the Pauli blocking method.

\begin{figure}[htbp]
\includegraphics[angle=0,scale=0.51]{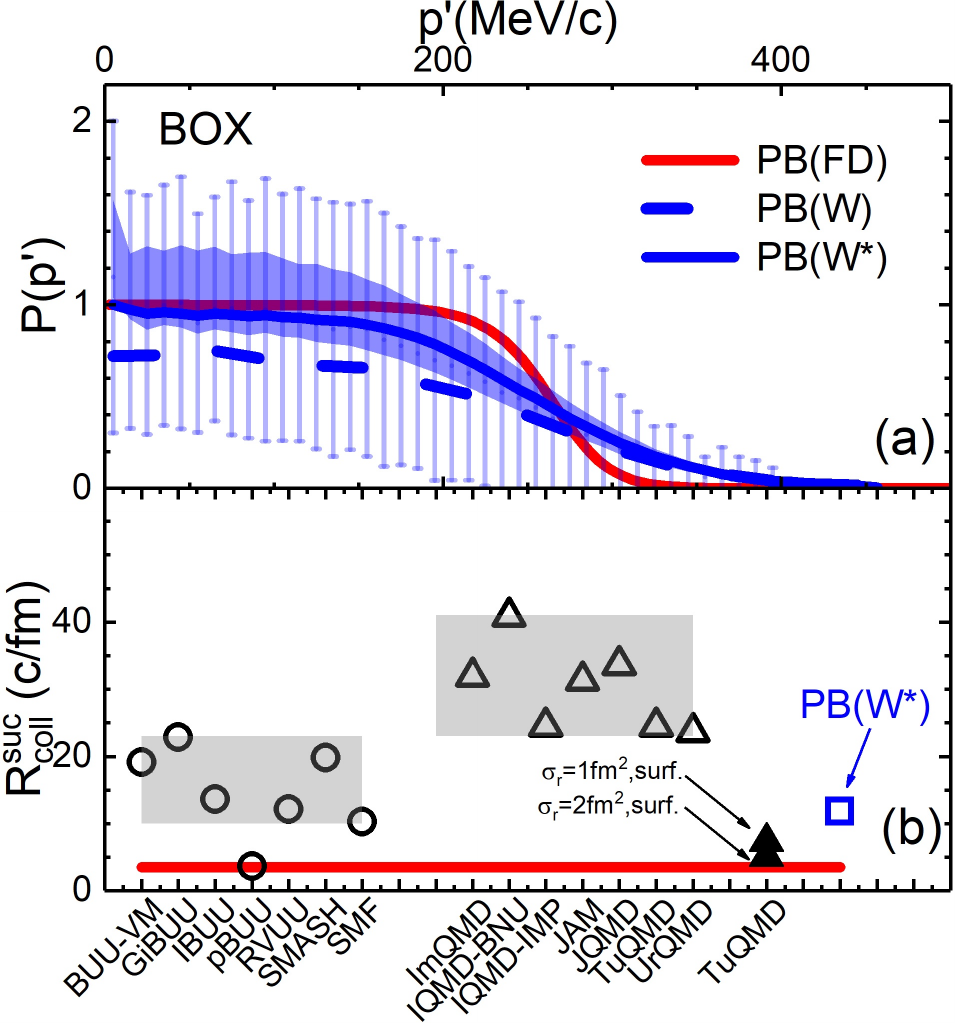}
\vspace{-0.9em}
\caption{(a) The standard deviation of $P(p')$ and the $\left<\mathrm{min} (P(p'),1) \right>$ by using the PB(W) and PB(W*). (b) The $R_{\mathrm{coll}}^{\mathrm{suc}}$ obtained in this work, and in Ref.~\cite{YXZhang2018PRC} for different transport models. 
The red line represents the benchmark values of the successful collision rate.} 
\label{fig:Fig2-BOX_AvOccPro}
\end{figure}

Figure~\ref{fig:Fig2-BOX_AvOccPro}(a) gives the distributions of the occupation probability for the final state of
all collisions in the first time step in nuclear matter at $T$=5 MeV. The conditions for simulations\footnote{the same number of nucleons (N=1280), the same width of the box ($L$=20 fm), the same cross section ($\sigma_{NN}=40$ mb), the same stop time ($t$= 140 fm/$c$), and the same number of events ($N_{event}=200$)} are also as the same as those in Ref.~\cite{YXZhang2018PRC}. The lines are the average values used in the practical calculations, i.e., $\langle \min{(P(p'),1)} \rangle$. Two cases are presented, one is for PB(W) (the blue dashed line with blue error bars), and another is for PB(W*) (the blue solid line with blue-shaded region). 
The benchmark result, which is a Fermi distribution at $T$=5 MeV, is presented as a red line. 
For the case of PB(W*), the variance of $P(p')$ is reduced by about 70\% compared to the conventional method. Moreover, the $\langle \min{(P(p'),1)} \rangle$ values obtained with PB(W*) are close to the benchmark values at low momentum region. At high kinetic energy region, the $\langle \min{(P(p'),1)} \rangle$ values obtained with PB(W*) are similar to those obtained with the conventional method and the $P(p')$ values are overestimated.

Figure~\ref{fig:Fig2-BOX_AvOccPro}(b) shows the averaged successful collision rate $R_{\mathrm{coll}}^{\mathrm{suc}}=\left< \mathrm{d}N_{\mathrm{coll}}^{\mathrm{suc}}/\mathrm{d}t \right>$, which is obtained during the time interval 60-140 fm/$c$. The open symbols are the results of the different BUU and QMD models, which have been published in Ref.~\cite{YXZhang2018PRC}. The blue square is the $R_{\mathrm{coll}}^{\mathrm{suc}}$ obtained with PB(W*), which is about 11.8 $c$/fm and comparable to the values obtained with IBUU, RVUU, and SMF models. In the QMD family, our results on $R_{\mathrm{coll}}^{\mathrm{suc}}$ evidence that the new Pauli blocking method is better than the conventional methods and worse than the hard sphere overlap method with surface correction in the TuQMD~\cite{Cozma2023}. The CoMD result is close to the benchmark value, but is not presented here since they use a different philosophy~\cite{Papa2001PRC,Papa2013PRC,Papa2005JCP}. However, the advantage of our method is that we did not introduce any new parameters which is different than the hard sphere overlap method with surface correction.

Next, we check the influence of PB(W*) on the finite nucleus which will be important for simulating HICs. As an example, we choose the $^{124} \mathrm{Sn}$ in the following analysis since it is on the long isotope chain and was widely used to study the isospin physics in HICs~\cite{Tsang2004PRL, Tsang2009PRL, Famiano2006PRL, Estee2021PRL,Jhang2021PLB,Lee2022EPJA}. Two quantities are analyzed. One is the distribution of the occupation probability, i.e., $P(p')$. Another is the radial distribution of the successful collision rate $\left< \mathrm{d}N_{\mathrm{coll}}^{\mathrm{suc}}/\mathrm{d}t\mathrm{d}r \right>$, which can help us identify the region where the spurious $NN$ collisions occur due to the defects of Pauli blocking. 
All the simulations are performed by using the SLy4 interaction, and the number of events is 1000. 

\begin{figure}[htbp]
\includegraphics[angle=0,scale=0.38]{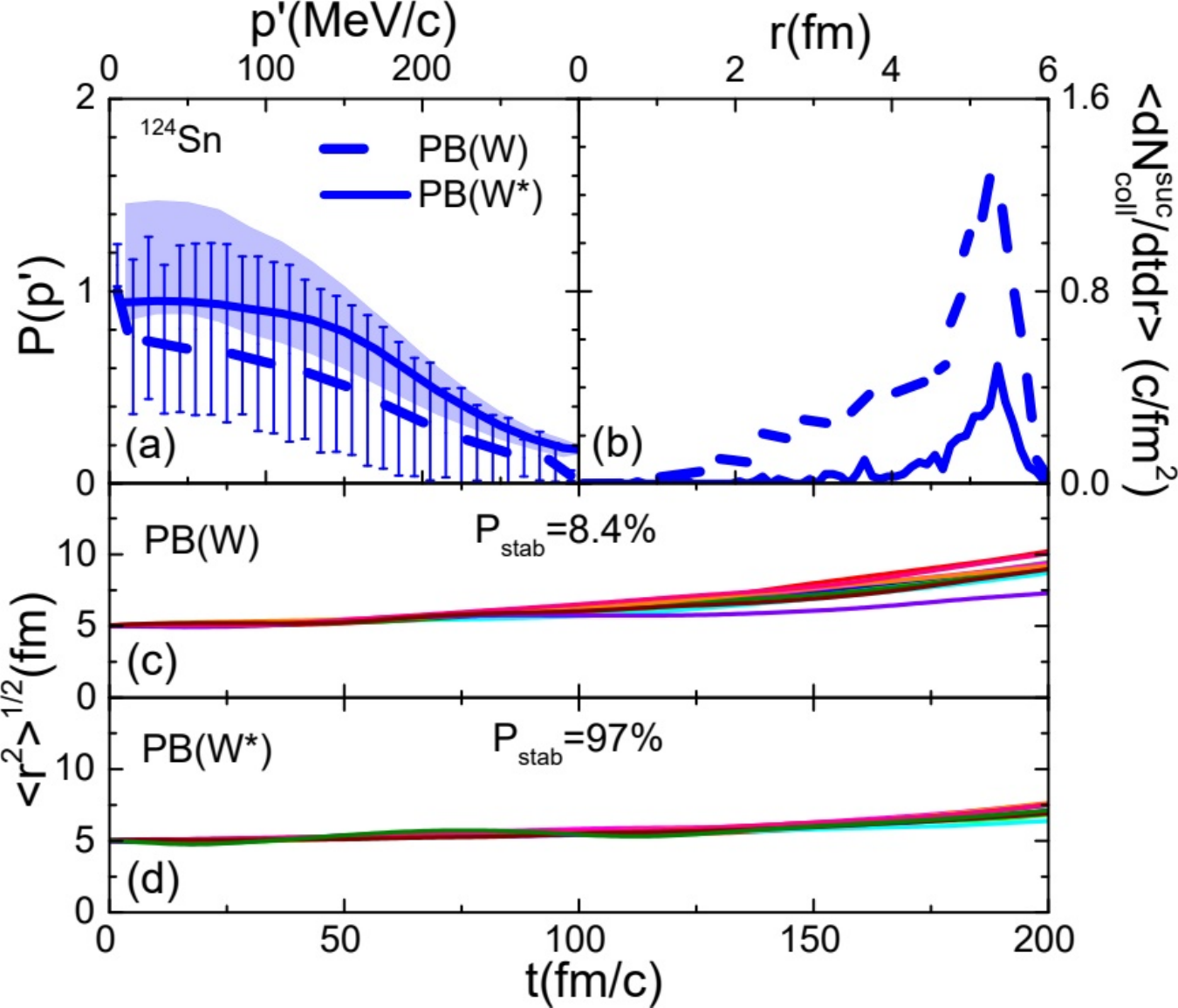}
\vspace{-0.9em}
\caption{(a) The standard deviation of $P(p')$ and the $\left<\mathrm{min} (P(p'),1) \right>$ by using the PB(W) and PB(W*). (b) The radial distribution of the successful collision rate for $^{124} \mathrm{Sn}$ for PB(W) and PB(W*). (c) and (d) Root-mean-square radius of $^{124} \mathrm{Sn}$ as a function of time for PB(W) and PB(W*), respectively. 
}
\label{fig:Fig3-Nucleus_AvOccPro}
\end{figure}

Figure~\ref{fig:Fig3-Nucleus_AvOccPro}(a) shows the distributions of the occupation probability for the initialized nucleus, i.e., $^{124}\mathrm{Sn}$, with PB(W) and PB(W*) at the first time step. Ideally, the momentum distribution of the initial nuclei $P(p')$ at the first time step still keeps the Fermi Dirac distribution. However, the shape of $P(p')$ obtained with both PB(W) and PB(W*) deviates from the expected Fermi distribution. This deviation can be attributed to two factors. First, the initial nucleus in the framework of QMD does not represent the true ground state, resulting in a binding energy that deviates by approximately 8\% from the true ground state~\cite{JPYang2021PRC}. Second, there is the defect of Pauli blocking as mentioned in Refs.~\cite{Aichelin1991PR,YXZhang2018PRC}. Comparing the results obtained with PB(W) and PB(W*), we find that the variance of $P(p')$ is reduced by about 35\% compared to the result by using the PB(W). Thus the averaged occupation probability, i.e., $\langle \min(P(p'),1) \rangle$, obtained with PB(W*) is greater than that for the PB(W). At low momentum regions, the occupation probability obtained with PB(W*) is close to 1. Panel (b) shows the radial distribution of the successful collision rate for PB(W) and PB(W*) at first time step. With PB(W*), the occurrence of spurious collision is suppressed by approximately 60\% in the surface region of the nucleus. 


In addition, we check the influence of the new Pauli blocking method on the stability of the initial nucleus, as the stability of the initial nucleus is crucial for the simulations of HICs. The root-mean-square (rms) radius of sampled $^{124} \mathrm{Sn}$ as a function of time is tested and plotted in Fig.~\ref{fig:Fig3-Nucleus_AvOccPro}(c) and (d) for PB(W) and PB(W*), respectively. The curves are from ten random events, and they illustrate that the new Pauli blocking method improves the stability of the initial nuclei. To quantify the stability of the sampled initial nucleus, we employ the probability of stability ($P_{\mathrm{stab}}$), as used in Ref.~\cite{JPYang2021PRC}. $P_{\mathrm{stab}}$ is calculated as the ratio of the number of events ($N_{\mathrm{stab}}$) that maintain the rms variation within 40\% at $t$=200 fm/$c$ compared to the rms values at $t$=0 fm/$c$, to the total number of events ($N_{\mathrm{total}}$), with $N_{\mathrm{total}}$ set to 1000. Our calculations reveal that the $P_{\mathrm{stab}}$ obtained with PB(W*) reaches 97\%, which is larger than that obtained with PB(W).



\textit{Impacts of new Pauli blocking method on HICs observables and the extraction of physical quantity.}
Now, let us turn to investigate the influence of the new Pauli blocking method on the HICs observables, such as the charge distributions, i.e., $dM/dZ$, and isospin sensitive observable, i.e., the free neutron to proton yield ratio, i.e., $R(n/p)$. 
The reaction system we simulated is $^{197}$Au+$^{197}$Au, and the beam energy is from 50 to 250 MeV/u. All the parameters we used in the simulations are the same except for the Pauli blocking method, allowing us to isolate the specific influence of the new method on the observables.

\begin{figure}[htbp]
\includegraphics[angle=0,scale=0.33]{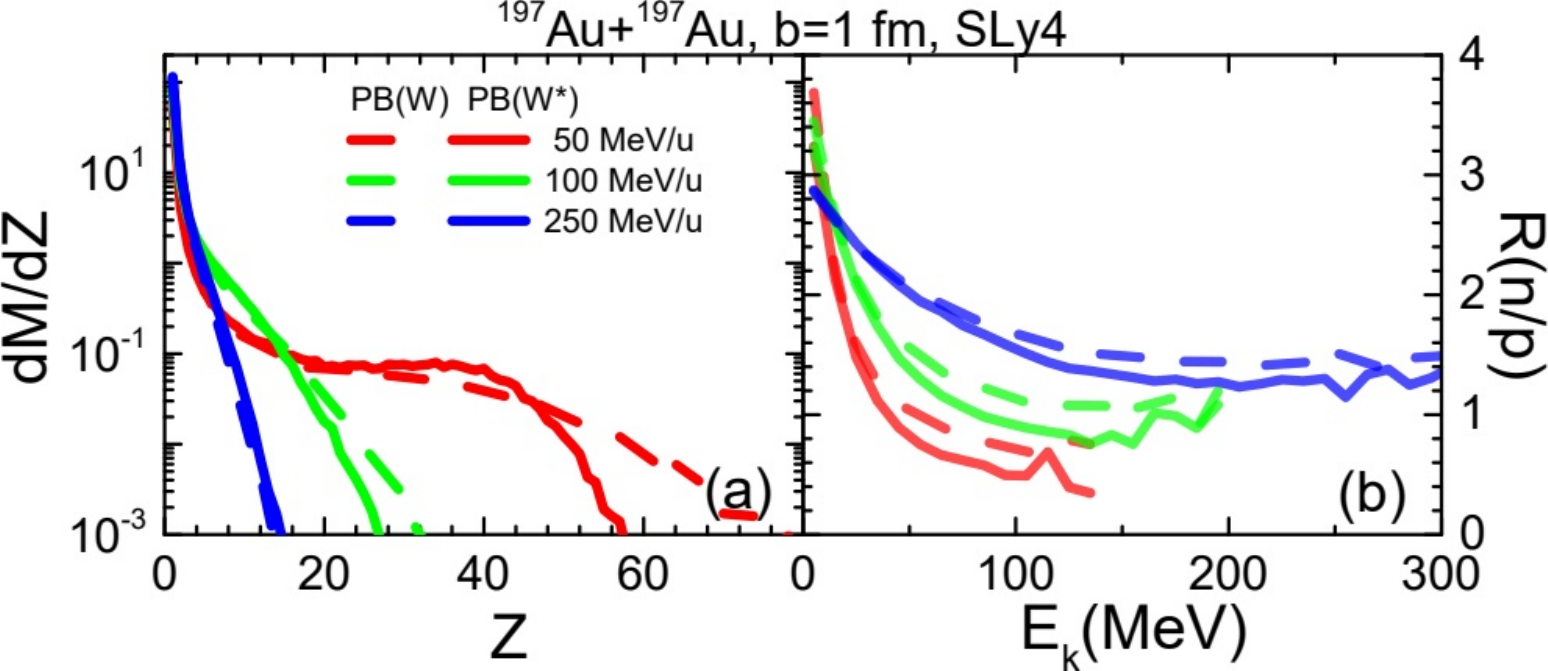}
\vspace{-1.0em}
\caption{(a) The charge distribution, and (b) The spectra of $R(n/p)$, obtained with PB(W) and PB(W*). }
\label{fig:Fig4-Z-Ratio}
\end{figure}

Figure~\ref{fig:Fig4-Z-Ratio}(a) shows the charge distributions of fragments obtained with PB(W) (dashed lines) and PB(W*) (solid lines). Our calculations show that the charge distributions obtained with the PB(W*) are narrower than those with PB(W), which can be attributed to the stability of the initial nuclei. As we learned in Fig.~\ref{fig:Fig3-Nucleus_AvOccPro}(c) and (d), the stability of the initial nuclei is better for PB(W*) than that for PB(W), which means that the projectile/target nuclei are tightly bound for PB(W*) than PB(W) during the time evolution. Consequently, the projectile and target are more likely to penetrate each other and form heavier fragments for PB(W) compared to PB(W*). With the beam energy increasing, the difference in the charge distribution obtained with two kinds of Pauli blocking becomes small as the weakening of the Pauli blocking effects.


Figure~\ref{fig:Fig4-Z-Ratio}(b) presents the influence of different Pauli blocking methods on the isospin sensitive observable, $R(n/p)$. Our results show that the calculations with PB(W*) lead to a suppression of the neutron to proton yield ratios compared to PB(W), especially at high kinetic energy of the emitted nucleons. For $R(n/p)$ ratios at $E_{beam}=50$ MeV/u, the suppression of $R(n/p)$ by using the PB(W*) is less than 15\% relative to that with PB(W). This suppression implies that the constraints of symmetry energy via transport models will modified if the PB(W*) is adopted, but the strength of the modification in the constraints of the symmetry energy depends on the correlation between the $R(n/p)$ and the slope of the symmetry energy.



To understand the impact on the extraction of physical quantities, as an example, we investigate the in-medium $NN$ cross sections. It was done by describing the stopping power $vartl$~\cite{Reisdorf2004PRL}, which measures the ratios between the variance of the transverse rapidity distributions and longitudinal distributions of the emitted particles and is closely related to the successful $NN$ collision rate. The $vartl$ is defined as 
\begin{equation}
  vartl=\frac{\langle y_{t}^{2}\rangle}{\langle y_{z}^{2}\rangle}.
\end{equation}
Here, $y_t$ and $y_z$ are the transverse and longitudinal rapidity of the detected particles. The calculations are performed for Au+Au at $b$=1 fm with the interaction parameter set SLy4. 

\begin{figure}[htbp]
\includegraphics[angle=0,scale=0.4]{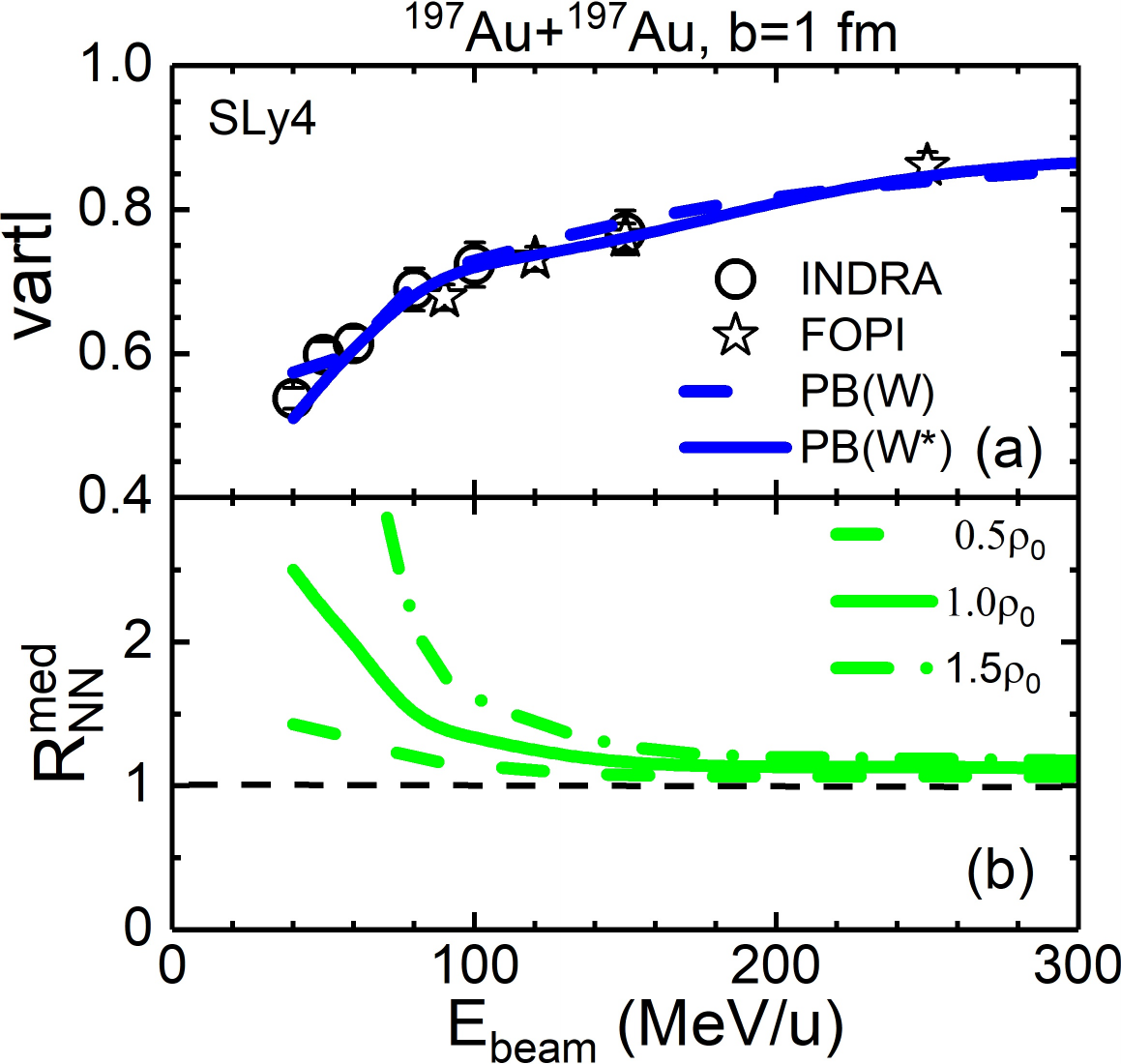}
\vspace{-1.0em}
\caption{(a) The excitation function of $vartl$ obtained with PB(W) and PB(W*). (b) The ratio of the extracted in-medium $NN$ cross section obtained with two Pauli blocking methods, line with different styles correspond to different medium density. The data are presented as symbols from Refs.~\cite{Andronic2006EPJA,Reisdorf2004PRL,Reisdorf2010NPA}.}
\label{fig:Fig4-vartl-eta}
\end{figure}

Figure~\ref{fig:Fig4-vartl-eta}(a) shows the excitation function of $vartl$ for Au+Au. The symbols are the data of $vartl$ obtained in Refs.~\cite{Andronic2006EPJA,Reisdorf2004PRL,Reisdorf2010NPA}. Two Pauli blocking methods have been adopted to describe the data, which are from INDRA and FOPI Collaborations~\cite{Andronic2006EPJA,Reisdorf2004PRL,Reisdorf2010NPA}. Panel (b) shows the ratio of the extracted in-medium $NN$ cross section between two methods, i.e, 
\begin{equation}
R_{NN}^{\mathrm{med}}=\frac{\sigma_{NN}^{\mathrm{med}}[PB(W^*)]}{\sigma_{NN}^{\mathrm{med}}[PB(W)]}.
\end{equation}
Three lines correspond to the results at $0.5\rho_0$, $\rho_0$, and $1.5\rho_0$. Our calculations show that the extracted $NN$ cross sections with the new Pauli blocking method are enhanced compared to the extracted in-medium $NN$ cross sections obtained with the conventional Pauli blocking method. At low incident energy, for example, at $E_{\mathrm{beam}}$ = 40 MeV/u, the extracted in-medium $NN$ cross sections at $\rho$=$\rho_0$ with PB(W*) are 2.5 times larger than that with PB(W). At the incident energy greater than 150 MeV/u, the values of $R_{NN}^{\mathrm{med}}$ tend to 1.1 at normal density. The enhancement of the in-medium $NN$ cross sections is similar as observed in the AMD calculations~\cite{Kaneko2021PLB}, but our results show that the cross sections extracted from PB(W) and PB(W*) are almost identical and previous results of the extracted in-medium $NN$ cross sections obtained with PB(W) do not need to change dramatically in the framework of the QMD model.


However, one should note that the above discussions are not sufficient to draw quantitative conclusions regarding the in-medium $NN$ cross sections. This is because we have not taken into account the impact parameter smearing effect, which exists in experiments, and the uncertainties arising from the effective interactions. A more comprehensive and accurate analysis of the in-medium $NN$ cross sections will be done in our next work. 

\textit{Summary and outlook.}
In this paper, we focus on proposing a novel method to calculate the occupation probability in the treatment of the Pauli blocking in the QMD-type models rather than to reproduce or explain the existing data. This method considers the quantum effect that each nucleon occupies a series of states rather than a single state, and it effectively reduces the fluctuations in the occupation probability and enhances the Pauli blocking ratios. The validation of the new Pauli blocking method has been done in the nuclear matter and the finite nuclei by comparing them with the benchmark values. The comparisons evidence that the new Pauli blocking method can provide a more accurate Pauli blocking ratio than the conventional Pauli blocking method used in QMD-type models.



Furthermore, we examine the influence of the new Pauli blocking method on the heavy ion collision observables, such as the charge distribution and the free neutron to proton yield ratios. Our calculations show that the influence of the Pauli blocking method on the charge distributions can not be neglected, especially at low beam energy, and the impact on the free neutron to proton yield ratios is less than 15\%. For the extracted physical quantity, such as the in-medium $NN$ cross sections, which are obtained by describing the experiment data of stopping power, our calculations show that the extracted in-medium $NN$ cross sections with the new Pauli blocking method are enhanced compared to that with the conventional Pauli blocking method. At the normal density, the extracted in-medium $NN$ cross sections with PB(W*) are enhanced 1.1$-$2.5 times larger than that with PB(W) at $E_{\mathrm{beam}}$=40$-$150 MeV/u. At $E_{\mathrm{beam}}>$150 MeV/u, the extracted in-medium $NN$ cross sections are enhanced 1.1 times larger than that with PB(W).


Overall, our results emphasize the importance of an appropriate Pauli blocking method in accurately describing heavy ion collisions and in improving insights into the reaction dynamics. The new Pauli blocking method provided in this work will be a better choice for the development of advanced transport models in the future.


\section*{Acknowledgements}

The authors thank Dr. Dan Cozma for providing useful suggestions and the number of successful collision rates obtained with the hard sphere overlap method + surface correction in the BOX calculations. This work was partly inspired by the transport model evaluation project, and it was supported by the National Natural Science Foundation of China under Grants No. 12275359, No. 12375129, No. 11875323 and No. 11961141003, by the National Key R\&D Program of China under Grant No. 2023 YFA1606402, by the Continuous Basic Scientific Research Project, by funding of the China Institute of Atomic Energy under Grants No. YZ222407001301 and No. YZ232604001601, and by the Leading Innovation Project of the CNNC under Grants No. LC192209000701 and No. LC202309000201. 

\bibliography{References}

\end{document}